\documentclass[a4paper,12pt]{article}
\usepackage{verbatim} 
\usepackage{jheppub} 
 \usepackage{amsmath,amssymb,braket,tikz}
\usetikzlibrary{tikzmark,calc}
\usepackage[T1]{fontenc} 
\usepackage{amsmath}
\usepackage[percent]{overpic}
\usepackage{wrapfig}
\usepackage{tabu}
\usepackage{diagbox}

\usepackage{graphicx}
\usepackage{tikz}
\usepackage{import}
\usepackage{accents}
\usepackage{mathrsfs,amsmath,amssymb,slashed}
\usepackage{multirow,multicol}
\usepackage[percent]{overpic}
\usepackage{slashed}

\usepackage{wrapfig}
\usepackage{tabu}
\usepackage{diagbox}
\usepackage{mathrsfs,amsmath,amssymb,amsthm,amsfonts,tikz,graphicx,accents,hyperref,color}
\usepackage{leftidx}
\usepackage{dsfont}
\usepackage{import}
\usetikzlibrary{decorations.pathmorphing}
\DeclareFontFamily{OT1}{rsfs}{}

\DeclareFontShape{OT1}{rsfs}{m}{n}{ <-7> rsfs5 <7-10> rsfs7 <10->rsfs10}{} 

\DeclareMathAlphabet{\mycal}{OT1}{rsfs}{m}{n}

\newcommand{\nn}{\nonumber}

\newcommand{\be}{\begin{equation}}
\newcommand{\ee}{\end{equation}}

\DeclareMathOperator{\extdm}{d}
\newcommand{\extd}{\extdm \!}

 \title{\bf 
  {The near horizon dynamics in three-dimensional Einstein gravity}}
  
\author[]{Hamid Afshar, Narges Aghamir}

\affiliation{\it Department of Physics, Faculty of Science, Ferdowsi University of Mashhad, Mashhad, Iran}
\emailAdd{ham.afshar@gmail.com, n.aghamir@gmail.com}

\abstract{ We study the asymptotic dynamics of 3D gravity with Rindler boundary conditions both in flat and AdS spacetimes. We do this by using the angular quantization and Hamiltonian reduction of the action to the Wess-Zumino-Witten theory on the boundary. 
We then rewrite the boundary action as a functional of elements of the asymptotic symmetry group.
}
\begin{document}
\maketitle
\section{Introduction}

The black hole near horizon physics has recently gained new attention in the framework of Jackiw-Teitelboim (JT) gravity \cite{Jackiw:1984je,Teitelboim:1983ux}.
JT gravity as a two-dimensional dilaton gravity describes the dynamics close to the horizon of near-extremal black holes. Under appropriate boundary conditions it provides a toy model for holography in the Rindler AdS$_2$ background, sometimes referred to as near AdS$_2$/near CFT$_1$ correspondence \cite{Kitaev:15ur,Kitaev:2017awl,Almheiri:2014cka,Maldacena:2016hyu,Maldacena:2016upp,Engelsoy:2016xyb,Cvetic:2016eiv,Jensen:2016pah}. Its holographic description is in terms of a one-dimensional Schwarzian action at the boundary for the Euclidean time reparametrization (which is the element of the asymptotic symmetry group). Similar holographic correspondences have been scrutinized in the 2D {\it flat} Rindler background (a near horizon description of non-extremal black holes) for  flat JT-models \cite{Callan:1992rs,Cangemi:1992bj,Afshar:2019axx,Godet:2020xpk,Afshar:2020dth,Godet:2021cdl,Afshar:2021qvi,Afshar:2022mkf}. The Euclidean boundary dynamics in all such cases is in terms of an effective 1D  Schwarzian-like action for elements of the asymptotic symmetry groups
\cite{Afshar:2019tvp,Afshar:2021qvi}.  


This paper aims to promote this study to three-dimensional gravity by considering the 3D Rindler boundary conditions in the framework of Einstein gravity proposed in \cite{Afshar:2015wjm}. We will obtain the 2D-induced theory at the boundary via Hamiltonian reduction for this set of boundary conditions. Ultimately, we derive a reparameterization theory for the non-extremal near-horizon geometry within 3D Euclidean Rindler spacetime, expressed in terms of elements of the asymptotic symmetry group. As the asymptotic symmetries suggest, this theory will be the geometric action for the warped Virasoro group.

\paragraph{Hamiltonian reduction.} The classical formulation of 3D pure gravity is equivalent to a Chern-Simons action \cite{Witten:1988hc,Achucarro:1986vz, Witten:2007kt}. It is well known that the Chern-Simons action after solving certain constraints reduces to Wess-Zumino-Witten (WZW) action on the boundary \cite{Witten:1988hf,Moore:1989yh, Salomonson:1989fw} which itself reduces to a  Liouville-like action under imposing specific boundary conditions \cite{Brown:1986nw,Coussaert:1995zp,Barnich:2013yka}. The Hamiltonian reduction from a  WZW-model to a Liouville-like theory was first applied by  \cite{Forgacs:1989ac,Alekseev:1988ce,Bershadsky:1989mf} in the case of SL$(2,{\mathbb R})$ gauge group. The resulting classical theory describes the effective action for a would-be holographic dual quantum field theory. Looking at this boundary theory from a different perspective, it should govern the dynamics of boundary gravitons which are the physical degrees of freedom from the gravitational point of view \cite{Cotler:2018zff}. 
Technically this corresponds to a path integral quantization of coadjoint orbits of the asymptotic symmetry group which in the case of the Virasoro group was first obtained by Alekseev and Shatashvili \cite{Alekseev:1988ce}. 
Their work showed that the classical phase space of smooth AdS$_3$ metrics continuously
connected to global AdS$_3$, is represented by (two copies of) an infinite dimensional quotient space  Diff$(S^1)/$PSL$(2;{\mathbb R})$ \cite{Barnich:2017jgw,Cotler:2018zff,Mertens:2018fds}, which admits a geometric quantization \cite{Witten:1987ty}.

\paragraph{Warped conformal symmetry.} Our project delves into the Alekseev-Shatashvili machinery within the context of warped conformal field theories. WCFT's are Lorentz violating 2D field theories characterized by Virasoro–Kac–Moody symmetries \cite{Hofman:2011zj, Detournay:2012pc, Afshar:2015wjm}.
These symmetries manifest in the near-horizon geometry of every extremal black hole across various spacetime dimensions, suggesting that WCFT could be a holographic counterpart to black hole solutions in flat, AdS, or dS spacetimes \cite{Aggarwal:2019iay,Aggarwal:2022xfd,Aggarwal:2023peg,Detournay:2023zni}. There are boundary conditions that yield a Kac-Moody-Virasoro algebra as their asymptotic symmetry algebra, such as CSS boundary conditions \cite{Compere:2013bya, Compere:2013aya} or Troessaert boundary conditions \cite{Troessaert:2013fma, Barnich:2014}, both applicable to AdS$_3$ space. We specifically employ the near horizon boundary conditions for asymptotically Rindler spacetime both in flat and AdS$_3$ which inherently possesses a horizon \cite{Afshar:2015wjm}.
If we manage to find the boundary action corresponding to it, it would be the 2D geometric action on the coadjoint orbit of the warped Virasoro group Diff$(S^1)\ltimes$C$^\infty(S^1)$ sometimes referred to as warped Schwarzian theory \cite{Afshar:2019tvp}.
This theory was previously studied in the holographic context for the Cangemi-Jakiwe theory \cite{Afshar:2019axx}, its supersymmetric version \cite{Afshar:2022mkf}, and its extension to spin-two dilaton-gravity \cite{Afshar:2020dth}.

The structure of the paper is as follows, in section \ref{sec2} after introducing the 3D Rindler gravity theory we offer a WZW formulation for the theory using angular quantization of the Chern-Simons formulation. Then via  Gauss decomposition of the gauge group, we obtain a Liouville-like boundary action for the flat Rindler boundary condition. In section \ref{sec3} we promote this theory to a reparametrization theory in terms of elements of the warped Virasoro group. In section \ref{sec4} we develop the same discipline for Rindler AdS$_3$ and finally we discuss various aspects of the holonomy conditions in section \ref{sec5}. 

\section{Rindler boundary conditions}\label{sec2}
The most famous boundary conditions in three dimensions are the Brown-Henneaux \cite{Brown:1986nw} and the Barnich-Comp\'ere \cite{Barnich:2006av} with maximally symmetric vacua in AdS and flat spacetimes correspondingly. Here in this work, we are interested in the Rindler-type boundary conditions which are not maximally symmetric. For instance, in the flat limit, we know that the Rindler vacuum is not  Poincar\'e invariant but is hyperbolic \cite{Melton:2023dee}. 

In three spacetime dimensions, we consider the following line element in the Bondi gauge,
 exhibiting  Rindler behavior near $r=0$ \cite{Afshar:2015wjm},
  \begin{align}\label{brbc}
     \extd s^2= -2\,a(u,x)r\extd u^2-2\extd u\extd r+\extd x^2+2\left(\eta(u,x)+2r/\ell\right)\extd x\extd u\,,
 \end{align}
 where $u$ is the retarded time. All coordinates range in $\mathbb{R}$. The 3D Einstein gravity equations $R_{\mu\nu}=-2/\ell^2\,g_{\mu\nu}$ enforce the free functions $a$ and $\eta$ to obey the following equations
 \begin{align}\label{eom67}
  \partial_x a=0\,,\qquad \partial_x\left(\partial_u \eta + a\, \eta\right)=0\,.
 \end{align}
Therefore, onshell the function $a$ depends solely on the retarded time $a=a(u)$ and $\eta$ is solved as
\begin{align}
 \eta=\eta(u)+k(x)e^{-\int^{u}\extd u' a(u')}\,.
\end{align}
We are fully off-shell unless stated explicitly in the rest of the paper.
We emphasize here that both functions $a$ and $\eta$ in our setup are state-dependent and thus are allowed to vary. This feature is qualitatively (and quantitatively) different from the alternative near-horizon boundary conditions \cite{Donnay:2015abr,Afshar:2016kjj,Afshar:2016uax,Donnay:2016iyk} and the ultimate near-horizon dynamics \cite{Grumiller:2019tyl} where Rindler acceleration function is held fixed.
As a consequence of imposing this set of boundary conditions, the asymptotic form of the metric \eqref{brbc} is preserved by  diffeomorphisms generated by a vector field $\xi$ producing a Lie derivative such that $\mathcal{L}_\xi g_{xr}={\mathcal L}_\xi g_{xx}={\mathcal L}_\xi g_{rr}={\mathcal L}_\xi g_{ru}=0$;
\begin{align}\label{asympKill}
    \xi^u=\epsilon(u)+\mathcal{O}(1/r)\,,\qquad\xi^x=\alpha(u)+\mathcal{O}(1/r)\,,\quad\xi^r=-r\dot\epsilon(u)+\mathcal{O}(1)\,,
\end{align}
where $\dot{}$ is derivative w.r.t. $u$.
The Lie bracket algebra of the infinite-dimensional allowed diffeomorphisms \eqref{asympKill} is the semi-direct sum of a Witt algebra and a U$(1)$ current algebra. 
 The central extension of this symmetry algebra is denoted as warped conformal symmetry which contains a Virasoro algebra and a U$(1)$ Kac-Moody algebra \cite{Hofman:2011zj,Detournay:2012pc,Afshar:2015wjm}. 
 \paragraph{Asymptotic region.} The Penrose diagram of the Rindler spacetime corresponding to the zero mode solution in \eqref{brbc} with $a=a_0$, $\eta=0$ and $\ell\to \infty$ is shown in Fig. 1. The Killing vector $\partial_u$ should be timelike, thus we have $g_{uu}=-2a_0r<0$ which says that we either have $(r>0,a_0>0)$ or $(r<0,a_0<0)$ corresponding to the regions I and III respectively. We thus have two asymptotic regions in general. However, there are different choices for placing the $u,r$ coordinates to generate a Rindler wedge. Here, instead of considering this pair of Rindler wedges, the setup we are considering is the Rindler wedge region I as the portion of the Penrose diagram of flat spacetime depicted in Fig. 2 covered by $r>0$ (and thus $a_0>0$). Thus we are dealing with only one asymptotic region $r\to \infty$. The ($u, r)$ coordinates in the region I can be viewed as describing the exterior of a finite temperature black hole accessible by the near horizon observer \cite{Maldacena:2016upp}.

 \begin{figure}[!h]
 	\begin{minipage}[c]{0.45\linewidth}
\centering 
\includegraphics[width=4.5cm]{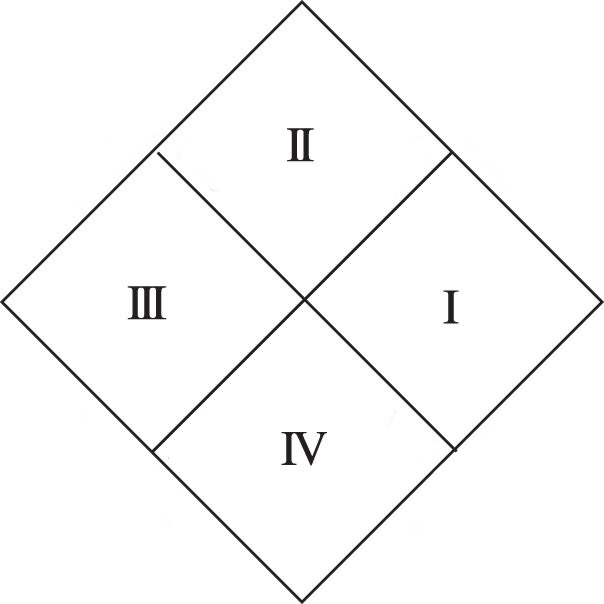} 
\caption{Penrose diagram of the Rindler spacetime.}	
\end{minipage}\hfill
\begin{minipage}[c]{0.5\linewidth}
	\centering
	\includegraphics[width=2.25cm]{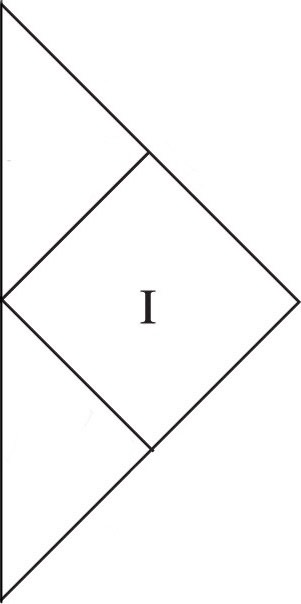}
	\caption{Rindler wedge portion of the Penrose diagram of flat spacetime.}
\end{minipage}
 \end{figure}
 \subsection{Chern-Simons formulation}
 The action that gives 3D Einstein gravity $I=\frac{1}{16\pi G}\int \extd^3x\sqrt{-g}(R \pm 2/\ell^2)$ is the Chern-Simons action up to a possible boundary term, for the gauge group ISO$(2,1)$, SO$(2,2)$ or SO$(3,1)$ depending on the value of the cosmological constant $\Lambda=\mp\frac{1}{\ell^2}$ \cite{Achucarro:1986vz,Witten:1988hc},
\begin{align}\label{CSAction}
    I_{{\text{\tiny CS}}}=\frac{k}{4\pi}\int_{\mathcal M}\left\langle A\wedge\extd A+\frac23A\wedge A\wedge A\right\rangle+I_{\text{\tiny bdy}}
\end{align}
where $k=1/(4G)$ and $A=e^a P_a+\omega^a J_a$ with $P$'s and $J$'s being generators of translation and Lorentz rotation. Here we consider the 3D manifold $\mathcal M=\Sigma\times\mathbb{R}$ with $\Sigma$ being the two-dimensional disk. The non-degenerate metric which should be used is the following
\begin{align}\label{algebra}
    \langle P_a\,J_b\rangle=\eta_{ab}\,.
\end{align}
The gauge algebra so$(2,2)$ can be written as
\begin{align}\label{so22algebra}
    [L_m,L_n]=(m-n)L_{m+n}\,,\quad[L_m,M_n]=(m-n)M_{m+n}\,,\quad[M_m,M_n]=\frac{1}{\ell^2} (m-n)L_{m+n}\,,\quad
\end{align}
where $L_{\pm1}=J_0\pm J_1$, $M_{\pm1}=P_0\pm P_1$ and $L_0=J_2$ and $M_0=P_2$ and thus $\langle L_1\,M_{-1}\rangle=\langle L_{-1}\,M_1\rangle=-2\langle L_0\,M_0\rangle=-2$.
We can recast our boundary conditions \eqref{brbc} in the first order form in terms of an algebra-valued connection,\footnote{We have made a finite gauge transformation with the SO$(2,2)$ group element $g=e^{\frac{r}{2}(M_{-1}-\frac1\ell L_{-1})}$ to eliminate the $r$-dependence.  All results will be valid in the flat limit $\ell\to \infty$ for the theory with gauge group ISO$(2,1)$.
} 
\begin{subequations}\label{bcnhfo}
\begin{align}
    A_u &=  M_1+a(u,x)L_0-\frac12(\dot\eta(u,x)+\eta(u,x)a(u,x))L_{-1} +\mathcal O(1/\ell)  \,,\\
    A_x&=M_0 -\frac12\eta(u,x)M_{-1}+\mathcal O(1/\ell)\,,\label{Axxx}\\
    A_r&=0\,,
\end{align}
\end{subequations}
where dot refers to derivative w.r.t. $u$. 
 The only boundary that we have is the one located in asymptotic infinity $r\to \infty$. The variation of the action made on-shell is,
\begin{align}\label{bdryaction}
          \delta  I_{\text{{\tiny CS}}}=-\frac{k}{4\pi}\int\extd^3 x \,\epsilon^{r\nu\rho}\partial_r\langle A_\nu  \delta A_\rho\rangle +\delta I_{\text{{\tiny bdy}}}\,.
       \end{align}
A well-defined variational principle for the first order action \eqref{arcsa} with the boundary conditions \eqref{bcnhfo} requires the addition of a boundary term of the form ($\epsilon^{rux}=1$) \cite{Afshar:2015wjm}
       \begin{align}\label{csawbt}
    I_{\text{\tiny bdy}}=-\frac{k}{4\pi}\int \extd u\extd x \,\langle A_u A_x\rangle\,.
\end{align}
Using the boundary condition \eqref{bcnhfo}, the boundary integrand is
\begin{align}
   \langle A_u A_x\rangle= a+\mathcal{O}(1/\ell)\,.
\end{align}

\subsection{From Chern-Simons to Wess-Zumino-Witten}

The stationary points of the Chern-Simons action are
 the flat connection $F=0$ and thus the theory has no local degrees of freedom and all dynamics are encoded in the holonomies \cite{Witten:1988hc} and boundary degrees of freedom \cite{Carlip:1994gy,Balachandran:1994up} which have been shown to reduce to Wess-Zumino-Witten (WZW) model on the boundary \cite{Elitzur:1989nr,Moore:1989yh}. Likewise, the 3D pure gravity with Brown-Henneaux  boundary conditions translates into a SO$(2,2)$ WZW model on the boundary which can be rewritten as a
Liouville theory \cite{Coussaert:1995zp}. 

The reduction of Chern-Simons theory to the WZW model which is based on the time foliation proceeds by solving the constraint $F_{\alpha\beta}=0$ at constant time where $\alpha$ and $\beta$ are spatial coordinates on the disk.
Our theory is specified by Rindler-type boundary conditions \eqref{brbc} where the geometry always has a horizon and the usual time foliation, breaks
down at the horizon. However, the vector field
$\partial_x$ is well defined everywhere, it is useful
to use this coordinate as time, and make a Hamiltonian
decomposition for the Chern-Simons action by slicing the Euclidean solid cylinder along the $x$-coordinate which means foliation by disks of constant $x$.
This is compatible with the idea of angular quantization in \cite{Banados:1998ys,Banados:2012ue}. This foliation is regular everywhere without any subtleties at the horizon. Making the $(2+1)$-decomposition:
\begin{align}
    A=A_\alpha \extd x^\alpha+A_x \extd x\,,
\end{align}
where $x^\alpha=(r,u)$ are coordinates on the $x$-constant slices, the Chern-Simons action \eqref{CSAction} becomes,
\begin{align}\label{arcsa}
    I_{\text{\tiny{CS}}}
    =\frac{k}{4\pi}\int \extd^3x\,\epsilon^{\alpha\beta}\left\langle  -A_\alpha \partial_xA_\beta+A_x F_{\alpha\beta} \right\rangle-\frac{k}{4\pi}\int\extd^3x\partial_r\langle A_uA_x\rangle+\frac{k}{4\pi}\int\extd^3x\partial_u\langle A_rA_x\rangle+I_{\text{\tiny bdy}}\,,
\end{align}
where $I_{\text{\tiny bdy}}$ is given in \eqref{bdryaction} and is equivalent with the first total $r$-derivative term appearing in \eqref{arcsa}, so they add up to $2I_{\text{\tiny bdy}}$. The total $u$-derivative term is zero for our boundary conditions.
Here by assuming that there are no nontrivial holonomies on the $x$-constant plane, a pure gauge configuration solves the constraint $F_{ru}=0$ by,
\begin{align}\label{sogf}
    A_\alpha=G^{-1}\partial_\alpha G\,, \qquad \alpha=r,u\,,
\end{align}
where $G$ is an arbitrary single valued group element.
A partial gauge fixing at constant $x$-slice is obtained by $\partial_u A_r=0$ which together with \eqref{sogf} induces the following factorization on $G$,
\begin{align}\label{gfc}
G(r,u,x)=g(x,u)b(r,x)\,.
\end{align}
This implies that,
\begin{align}
A_r=b^{-1}\partial_rb\,,\qquad A_u=b^{-1}A(u,x)b\,,
\end{align}
where $A(u,x)=g^{-1}\partial_u g\equiv g^{-1}\dot g$.
Substituting the above solution into the bulk action \eqref{arcsa} we have\footnote{We take $\epsilon^{ur}=-1$.}
\begin{align}\label{CSgb}
   I_{\text{\tiny CS}}&= \frac{k}{4\pi}\int \extd^3x \, \langle A_uA_r'-A_rA_u' \rangle+2I_{\text{\tiny bdy}}\nn\\
   &=\frac{k}{4\pi}\int \extd^3x  \,\langle -b^{-1}Ab'b^{-1}\partial_rb+b^{-1}A\partial_rb'+\partial_rbb^{-1}b'b^{-1}A-\partial_rbb^{-1}A'-b^{-1}\partial_rbb^{-1}A b'\rangle+2 I_{\text{\tiny bdy}}\,.
\end{align}
On the other hand, the Wess-Zumino term for the group element $G$ has the following form  $\epsilon^{rux}=1$,
\begin{align}\label{WZterm}
 \Gamma[G]&=\frac13\int\extd^3x\, \epsilon^{\mu\nu\rho}\langle G^{-1}\partial_\mu G G^{-1}\partial_\nu G G^{-1}\partial_\rho G\rangle\nn\\
 &=\int  \extd^3x \,\langle -\partial_rbb^{-1}g^{-1}g'A+ \partial_rbb^{-1}A g^{-1}g'+b^{-1}\partial_rbb^{-1}Ab'-\partial_rbb^{-1}b'b^{-1}A\rangle,
\end{align}
where we plugged in  $A$ for $g^{-1}\dot g$.
If we partially integrate the term containing $A'$ in \eqref{CSgb} over the $u$-coordinate the Chern-Simons action \eqref{CSgb} takes the following form,
\begin{align}\label{RWZW2}
    I_{\text{\tiny CS}}&=\frac{k}{4\pi}\left(-\Gamma[G]+\int \extd^3x\,\partial_r \langle b^{-1}A b' \rangle-\int \extd^3x\,\partial_u\langle\partial_r b b^{-1}g^{-1}g'\rangle\right)+2I_{\text{\tiny bdy}}\,.
\end{align}
The last term in the parenthesis of \eqref{RWZW2} is zero if we take $u$ to be periodic,
\begin{align}\label{uperiodic}
 u\sim u+2\pi L\,.
\end{align}
This choice is naturally suggested by the asymptotic symmetry analysis of \cite{Afshar:2015wjm} which led to a warped-witt algebra at vanishing $u$(1) level. 
 Although we take the retarded time to be periodic, we believe our final result for the effective action describing the reparametrization mode in any two-dimensional warped CFT holds independently of this assumption.
The integrand of the second term in \eqref{RWZW2} can be written covariantly as a sigma model term $\partial_r\langle G^{-1}\dot G G^{-1}G'\rangle$ which is a surface term. Now if we assume $b'=0$ on the boundary, the final form of the action is  a WZW action for the group element $G$,
\begin{align}\label{RWZW}
    I_{\text{\tiny CS}}&=\frac{k}{4\pi}\left(-\Gamma[G]+\int_{\partial\mathcal M}\langle g^{-1}\dot g\, g^{-1}g' \rangle\right)+2I_{\text{\tiny bdy}}\,.
\end{align}
This shows that the asymptotic dynamics of the $x$-foliated 3D gravity are similar to the time-foliated theory 
 is also described by the WZW action. We notice the crucial relative sign between the WZ and the sigma-model terms in \eqref{RWZW} which is indeed a direct consequence of the $x$-foliation of the theory. Had we used the time foliation in the CS action, the relative sign between these two terms would have been reversed. 
\paragraph{Gauss decomposition.}
It is useful to express the WZW
action \eqref{RWZW} in local form upon performing a Gauss decomposition of the group element $G$. We now decompose each  group elements $G(u;x; r)$ as \cite{Zhelobenko,Merbis:2019wgk},
\begin{align}\label{GDPSN}
	G=e^{XL_1} e^{WM_1} e^{\Phi L_0} e^{\zeta M_0} e^{YL_{-1}} e^{VM_{-1}}\,,
\end{align}
where $(X,\zeta,Y,W,\Phi,V)$ are functions of $(u, x, r)$ and their pull-back to the boundary, $g$, depends
only on $(u, x)$.
Here in this section, we consider $\ell\to\infty$ corresponding to flat Rindler boundary condition where the gauge group is identified as ISL$(2, \mathbb{R})$ --- for similar analysis with other boundary conditions in the 3D flat spacetime see
\cite{Salomonson:1989fw,Barnich:2012aw,Barnich:2013yka,Merbis:2019wgk}.
 Using the Gauss decomposition \eqref{GDPSN}, we rewrite the  Sigma model term in the action \eqref{RWZW},
\begin{align}
   \int_{\partial\mathcal M}\langle g^{-1}\dot g \,g^{-1} g' \rangle&=-2\int\extd u \extd x\Big [ e^\Phi \left ( \dot VX'+ \dot XV'+\dot YW'+\dot WY'+\zeta(\dot XY'+\dot YX')\right )\nn\\&\qquad\qquad\qquad-\frac12(\dot\Phi \zeta'+\dot \zeta \Phi')\Big ]\,.
\end{align}
The Wess-Zumino term in \eqref{RWZW} under the decomposition \eqref{GDPSN} can be written as a total $r$-derivative,
\begin{align}
     \Gamma[G]=-2\int \extd^3x \,\partial_r\Big [ e^\Phi \left ( \dot VX'- \dot XV'+\dot YW'-\dot WY'+\zeta(\dot YX'-\dot XY')\right ) \Big ]\,.
\end{align}
The final form of the $x$-foliated reduced CS action \eqref{RWZW} by the above Gauss decomposition reads as
\begin{align}\label{WZWunimprvd}
    I_{{\text{\tiny CS}}}=-\frac{k}{\pi}\int \extd u \extd x \Big [ e^\Phi \left ( \dot X( V'+\zeta Y')+\dot WY'\right ) -\frac14(\dot\Phi \zeta'+\dot \zeta \Phi')\Big ]+ 2I_{{\text{\tiny bdy}}}\,.
\end{align}
We stress again the fact that, in the time foliation of the CS action, the relative sign between the WZ and the sigma model terms in the WZW action \eqref{RWZW} are reversed and, as a consequence, the final reduced CS action in terms of the Gauss decomposed fields is completely different.

The solution \eqref{sogf} can  be represented in terms of the Gauss decomposed fields in \eqref{GDPSN} for ISL$(2,\mathbb{R}$) as,
 \begin{align}\label{grgf}
   A_u= G^{-1}\dot G&=\left[\dot{\Phi}+2e^{{\Phi}}Y\dot X\right]L_0+\left[{\dot \zeta}+2e^{{\Phi}}(V\dot X+Y\dot W+\dot XY\zeta)\right]M_0\nn\\&+\left[\dot Xe^{{\Phi}}\right]L_1+\left[\dot W+\dot X\zeta\right]e^{{\Phi}}M_1+\left[Y\dot \Phi+\dot Y+e^{{\Phi}}Y^2\dot X\right]L_{-1}\nn\\&+\left[Y\dot \zeta+V\dot \Phi+\dot V+e^{{\Phi}}(2YV\dot X+Y^2\dot W+\zeta Y^2\dot X)\right]M_{-1}\,.
 \end{align}
Comparing the boundary conditions \eqref{bcnhfo} with the gauge field components $(L_1,M_1,M_0,M_{-1})$ in \eqref{grgf} we are led to the following constraints among the Gauss decomposed fields,
\begin{align}\label{gc}
    \dot X=0\,,\qquad  \dot W=e^{-\Phi}-\dot X\zeta\,,\qquad \dot\zeta=-2Y-2V\dot X  e^{\Phi}\,,\quad  \dot V=-\dot\Phi V+Y^2\,.
\end{align}
Identification along $L_0$ and $L_{-1}$ determines the state-dependent functions $a(u,x)$ and $\eta(u,x)$ via equations 
\begin{align}\label{etazeta}
a=\dot\Phi+2Y\dot Xe^{\Phi}\,,\qquad
-\frac12(\dot\eta+2\dot Y)-\frac12(\eta+2Y)+Y^2e^{\Phi}\dot X=0\,,
\end{align}
which upon imposing $\dot X=0$ leads to $a=\dot\Phi$ and $\eta=\dot\zeta$.

Plugging in the constraints \eqref{gc}
into \eqref{WZWunimprvd} we have the following boundary action,
\begin{align}\label{wzwf}
    I_{\text{\tiny RFlat}}[\Phi,\zeta]
    =\frac{k}{4\pi}\int\extd u \extd x \left (\dot\Phi \zeta'+\dot \zeta \Phi' -2\dot\Phi\right)\,,
\end{align}
 where we assumed the fields are zero at $x=\pm\infty$ or have the same value at endpoints, such that
the total $x$-derivative terms have vanished.
 This action is a Liouville-like theory for the near horizon flat boundary conditions \eqref{brbc}. For comparison to the BMS$_3$ Liouville theory, we refer to \cite{Barnich:2012rz,Barnich:2013yka,Merbis:2019wgk}.

We emphasize that \eqref{wzwf} is the boundary action obtained upon reducing gravity in region I (see fig. 2) to its sole right asymptotic boundary as a consequence of restricting ourselves to the right Rindler wedge as explained in the beginning of section \ref{sec2}. This choice is valid for us because our goal is ultimately to use analytic continuation to translate our asymptotic dynamics in favor of the Euclidean theory in section \ref{sec5} which possesses only one boundary. In general, if one considers both asymptotic regions in the Lorentzian theory, reducing the gravity theory to boundary dynamics is more complicated as one should include Wilson lines stretched between left and right boundaries. For consulting details of this line of analysis follow \cite{Harlow:2018tqv,Cotler:2020ugk,Henneaux:2019sjx}.
 \section{Reparametrization theory}\label{sec3}
 Although the two-dimensional Liouville-like theory \eqref{wzwf} can be regarded as the effective 2D field theory describing the dynamics of the bulk theory, this description is not unique. It depends on the choices of the Gauss decomposition. In the rest, we try to fix this ambiguity by finding a field redefinition that represents the boundary action \eqref{wzwf} in terms of elements of the asymptotic symmetry group of the theory. In a more technical term, it is a geometric action on coadjoint orbits of the asymptotic symmetry group of the theory. It is thus useful to recall this group of symmetries from \cite{Afshar:2015wjm} before we continue our discussion about the boundary action. The allowed infinitesimal diffeomorphism \eqref{asympKill} that preserves our boundary conditions either in metric formulation \eqref{brbc} or its gauge field formulation \eqref{bcnhfo} act  on the boundary as
\begin{align}\label{2Ddiffeo}
	u\to u+\epsilon(u)\,,\qquad x\to x+\alpha(u)\,.
\end{align} 
The finite form of the action \eqref{2Ddiffeo} is the diffeomorphism $u\to f(u)$ and the supertranslation $x\to x+\alpha\circ f(u)$. 
The pair $(f,\alpha)$  are elements of the warped-Virasoro group Diff$(S^1)\ltimes$C$^\infty(S^1)$. 
As finite 2D diffeo's, they have the following periodicity along the closed path $u$,
\begin{align}\label{2Dfin}
	f(u+2\pi L)=f(u)+2\pi L\,,\qquad \alpha\circ f(u+2\pi L)=\alpha\circ f (u)\,.
\end{align}
The state-dependent functions $(a,\eta)$ transform infinitesimally under these diffeomorphism as
\begin{align}\label{asyma}
\delta a= \dot\epsilon a+\epsilon\dot a-\ddot\epsilon\,,\qquad\delta\eta=\dot\epsilon\eta+\epsilon\dot\eta+\dot\alpha\,.
\end{align}
The finite form of this set of transformations is the following \cite{Afshar:2015wjm}
\begin{align}\label{finitetrans}
	\tilde a\circ f=\frac{1}{\dot f}\left(a+\frac{\ddot f}{\dot f}\right)\,,\qquad	\tilde \eta\circ f=\frac{1}{\dot f}\eta-\alpha'\circ f \,,
\end{align}
where $\alpha'\circ f=\partial_f\alpha$.  It is easy to verify that upon taking $f(u)=u+\epsilon(u)$, the transformation \eqref{finitetrans} reduces to \eqref{asyma} infinitesimally.
In \eqref{finitetrans}  the fields $a$ and $\eta$ are mapped to a new configuration  $\tilde a$ and $\tilde \eta$ under the finite symmetry transformation ($f,\eta$). 
The equation \eqref{finitetrans} thus defines the  space of allowed  metric fluctuations around the background   configuration ($a_0,\eta_0$) 
 \begin{align}\label{constrepre}
     a(u)=a_0 \,\dot h-\frac{\ddot h}{\dot h}\,,\qquad\eta(u)=\eta_0\,\dot h-\dot\alpha\,,
 \end{align}
where $h=f^{-1}$ is a general diffeomorphism on $S^1$.

 The boundary action \eqref{wzwf}  should be found as a functional of ($h,\alpha$) on the background configuration ($a_0,\eta_0$). We thus need to know the dependence of the functions $(\Phi,\zeta)$ appearing in \eqref{wzwf} to both the background values ($a_0,\eta_0$) and to the fields $(h,\alpha)$. By studying their periodicity properties, we fix the dependence of the fields appearing in the group element $g$ to ($h,\eta$).  
Using the constraints imposed via boundary conditions, $a=\dot\Phi$ and $\eta=\dot\zeta$ we get 
\begin{align}\label{ident}
	\Phi=a_0\, h -\log \dot h +B\,,\qquad\zeta=\eta_0 \,h -\alpha+D\,,
\end{align}
with the following periodicity property induced from \eqref{2Dfin}  (at fixed $x$)
\begin{align}\label{holfc}
	\Phi(u+2\pi L)&=\Phi(u)+(2\pi L)a_0\,,\qquad\zeta(u+2\pi L)=\zeta(u)+(2\pi L)\eta_0\,.
\end{align}
The $u$-independent functions $B(x)$ and $D(x)$ are integration constants we set to zero for now.
After applying the field redefinition \eqref{ident}, and promoting the $h(u)$ and $\alpha(u)$ to functions of $x$, the form of our boundary action \eqref{wzwf} as a functional of $h(u,x)$ and $\alpha(u,x)$ on a given background $(a_0,\eta_0)$   is 
\begin{align}\label{boundaryaction}
    I[h,\alpha]
        &=\frac{k}{4\pi}\int \extd u\extd x \Big[2a_0\eta_0\, \dot hh'-a_0(\dot h\alpha'+\dot\alpha h')-\eta_0\left(h'\frac{\ddot h}{\dot h}+\dot h'\right)+\dot\alpha\frac{\dot h'}{\dot h}+\alpha'\frac{\ddot h}{\dot h}-2a_0 \dot h+2\frac{\ddot h}{\dot h}\Big]\,.
\end{align}
The field equations derived from this action are
\begin{align} 
   &\partial_u\left(a_0h-\log \dot h\right)'=0\,,\\ 
&\partial_u\left(-2a_0\eta_0h'+a_0\alpha'+\frac{\dot\alpha'}{\dot h}\right)=0\,.
\end{align}
It is easy to verify that these equations reproduce equations  \eqref{eom67}.  
The action \eqref{boundaryaction} can be considered as the two-dimensional warped-Schwarzian theory when the U(1) Kac-Moody level is zero. The one-dimensional theory was introduced in \cite{Afshar:2019tvp} as the group action for the twisted warped Virasoro coadjoint orbits on the circle. The vanishing level case is especially interesting because one can not remove the mixing term between $\alpha$ and $h$ via field redefinition. The same happens in the warped Virasoro symmetry algebra;
\begin{align}
&[L_n,L_m]=(n-m)L_{n+m}+\frac{c}{12}n^3\delta_{m+n,0}\,,\nn\\
&[L_n,P_m]=-mP_{n+m}+\kappa n^2\delta_{m+n,0}\,,\nn\\
&[P_n,P_m]=\frac{K}{2}n\delta_{m+n,0}\,,
\end{align}
 when the U(1)-level $K$ is zero and thus the central charge between generators of $S^1$-diffeo's $L_n$ and supertranslaiton sector $P_n$ can not be twisted away. Below we consider the case where twisting is possible ($K\neq0$) by going to the Rindler AdS case where $K\propto1/\ell$ and thus the circle reparametrization modes and the supertranslation modes can potentially be decoupled.
\section{Boosted Rindler-AdS}\label{sec4}
If we add a cosmological constant $\Lambda=-1/\ell^2$ to the theory, the $xu$ component of the metric acquires a term proportional to the radial coordinate as in \eqref{brbc}.   We can write the line element in first-order formalism in terms of the dreibein as,
\begin{align}
	ds^2=\eta_{ab}e^ae^b=-4e^1e^{-1}+(e^0)^2\,.
\end{align}
So we have,
\begin{align}
	e^0=dx\,,\qquad e^1=\extd u\,,\qquad e^{-1}=\frac{1}{2} \left ( {a(u)r}\extd u-(\eta(u)+\frac{2r}{\ell})\extd x+\extd r \right )\,.
\end{align}
Using the torsion  equation, we can drive the (dualized) spin connection components as,
\begin{align}
	\omega^1=\frac{1}{\ell}\extd u \,,\quad \omega^0=a(u)\extd u-\frac{1}{\ell}\extd x\,,\quad\omega^{-1}=\frac12\left\{  -(\eta'+a\eta{ +\frac{r}{\ell}a}) \extd u+\frac{1}{\ell}(\eta+\frac{ 2r}{\ell})\extd x-\frac{1}{\ell}\extd r\right\}\,.
\end{align}
We can remove the $r$-dependence  by a gauge transformation  using the following group element,
\begin{align}
	b=\text{exp}[\frac{r}{2}(M_{-1}-\frac{1}{\ell}L_{-1})]\,.
\end{align}
So we have,
\begin{align}\label{rads}
	A^{\text{\tiny RAdS}}=A^{\text{\tiny Rflat}}+\frac{1}{\ell}\left(-\extd x\,L_0+\frac12\eta \extd x \,L_{-1}+\extd u\,L_1\right)\,,
\end{align}
where $A^{\text{\tiny Rflat}}$ is given in \eqref{bcnhfo} for $\ell\to\infty$.
In our parametrization of the algebra, the isl$(2,\mathbb R)$  deforms to the so$(2,2)$ algebra \eqref{so22algebra} for finite $\ell$. 
For the asymptotic AdS space, we use the same Gauss decomposition as the flat case \eqref{GDPSN}, 
but here we take care of the added commutator, $[ M_n,M_m  ]=\frac{1}{\ell^2}(n-m)L_{m+n}$. 
As in the previous case, we assume $2\pi L$-periodicity in $u$. Again, assuming no nontrivial holonomies on the $x$-constant disc, we solve the constraint $F_{ru}=0$ by, $A_u=G^{-1}\dot G$ where $G$ is the arbitrary single valued group element of SO$(2,2)$ which is decomposed according to \eqref{GDPSN}. A precise calculation without the use of any specific representation for the group element leads to, 
\begin{align} \label{grgfads}
	G^{-1}\dot G&=\Big[e^\Phi\Big (\text{cosh} (\frac{\zeta}{\ell})\dot X+\frac{1}{\ell}\text{sinh} (\frac{\zeta}{\ell})\dot  W\Big )\Big]L_1+\Big[e^\Phi\Big ( \ell \text{sinh} (\frac{\zeta}{\ell})\dot X+\text{cosh} (\frac{\zeta}{\ell})\dot W\Big )  \Big]M_1\nn\\&+\Big [ 2e^\Phi\Big (Y \text{cosh} (\frac{\zeta}{\ell})+\frac{V}{\ell}\text{sinh} (\frac{\zeta}{\ell}) \Big )\dot X+2{e^\Phi}\Big ( \frac{Y}{\ell}\text{sinh} (\frac{\zeta}{\ell})+\frac{V}{\ell^2}\text{cosh} (\frac{\zeta}{\ell})\Big )\dot  W+ \dot \Phi\Big ]L_0\nn\\&+ \Big [ 2e^\Phi\Big ( V\text{cosh} (\frac{\zeta}{\ell})+Y\ell \text{sinh} (\frac{\zeta}{\ell})\Big )\dot X+ 2e^\Phi\Big (\frac{V}{\ell}\text{sinh} (\frac{\zeta}{\ell})+Y\text{cosh} (\frac{\zeta}{\ell})\Big )\dot W+\dot\zeta \Big ]M_0\nn\\&+\Big [e^\Phi\Big ((Y^2+\frac{V^2}{\ell^2}) \text{cosh} (\frac{\zeta}{\ell})+2\frac{VY}{\ell}\text{sinh} (\frac{\zeta}{\ell}) \Big )\dot X+{e^\Phi}\Big ( (\frac{Y^2}{\ell}+\frac{V^2}{\ell^3})\text{sinh} (\frac{\zeta}{\ell})+2\frac{VY}{\ell^2}\text{cosh} (\frac{\zeta}{\ell})\Big )\dot  W\nn\\&\qquad+Y\dot  \Phi+\frac{V}{\ell^2}\dot  \zeta+\dot  Y\Big ]L_{-1}\nn\\&+\Big [e^\Phi\Big ((\ell Y^2+\frac{V^2}{\ell}) \text{sinh} (\frac{\zeta}{\ell}) +2{VY}\text{cosh} (\frac{\zeta}{\ell})\Big )\dot X+{e^\Phi}\Big ( ({Y^2}+\frac{V^2}{\ell^2})\text{cosh} (\frac{\zeta}{\ell})+\frac{2VY}{\ell}\text{sinh} (\frac{\zeta}{\ell})\Big )\dot  W\nn\\&\qquad+Y\dot  \zeta+{V}\dot  \Phi+\dot  V\Big]M_{-1}\,.
\end{align}
It is easy to check that the above expression reduces to \eqref{grgf} in the limit $\ell\to\infty$. 
The WZW action \eqref{RWZW} can now be written for the SO$(2,2)$ group element $G$. We calculate the WZ term \eqref{WZterm} in the action \eqref{RWZW} by
substituting \eqref{grgfads} and find it as a total derivative,
\begin{align}
	\Gamma[G]=&\frac23 \int \extd^3x\,\epsilon^{\mu\nu\rho} \partial_\mu\Big(e^{\Phi}(\ell \partial_\nu X\partial_\rho Y+\frac{1}{\ell}\partial_\nu W\partial_\rho V)\text{sinh} \frac{\zeta}{\ell}+e^{\Phi}(\partial_\nu W\partial_\rho Y+\partial_\nu X\partial_\rho V)\text{cosh} \frac{\zeta}{\ell}\Big)\,.
 \nn\\
\end{align}
The sigma model term in the action \eqref{RWZW} after substituting the so$(2,2)$ valued gauge field \eqref{grgfads} takes the form, 
\begin{align}
	\int_{\partial{\mathcal M}}\langle g^{-1}\dot g\, g^{-1}g' \rangle=&-2\int \extd u \extd x\, e^{\Phi}\Big [ (\dot XV'+\dot V X'+\dot W Y'+\dot YW')\text{cosh} (\frac{\zeta}{\ell})\nn\\& +(\ell\dot XY'+\ell\dot Y X'+\frac{1}{\ell}\dot WV'+\frac{1}{\ell}\dot VW')\text{sinh} (\frac{\zeta}{\ell})-\frac{1}{2}(\dot \Phi\zeta'+\dot \zeta \Phi')\Big ]\,.
\end{align}
Finally the total action \eqref{RWZW} in the AdS case takes the form,
\begin{align}\label{csactionads}
	I_{{\text{\tiny CS}}}=&-\frac{k}{\pi}\int \extd u \extd x\, \Big [ e^{\Phi}\Big (  (\dot X V'+\dot WY')\text{cosh} (\frac{\zeta}{\ell})+(\ell\dot XY'+\frac{\dot WV'}{\ell})\text{sinh} (\frac{\zeta}{\ell})\Big)-\frac{1}{4}(\dot \Phi\zeta'+\dot \zeta \Phi')+\frac12\,a\Big]\,,
\end{align}
where in the last term we evaluated the boundary action \eqref{csawbt} using the fact that for boundary condition \eqref{rads} we have $\left<A_u\,A_x \right>=a$. 
Again the action \eqref{csactionads} agrees with \eqref{WZWunimprvd} after taking the $\ell\to\infty$ limit.

Adapting \eqref{grgfads} along $(L_1,M_1,M_0,M_{-1})$ to the gauge field $A_u$ introduced by the boundary condition \eqref{rads}, we find,
\begin{align}\label{constrainteqs}
	&\dot X e^\Phi=\frac{1}{\ell}e^{-\zeta/\ell}\,,\qquad \dot W e^\Phi=e^{-\zeta/\ell}\,,\\& \frac{2V}{\ell} + 2Y + \dot \zeta = 0\,,\qquad \frac{V^2}{\ell^2} + \dot V + \frac{2 V Y}{\ell} + Y (Y + \dot \zeta) + V \dot\Phi=0\,.
\end{align}
Substituting the above constraints into the action \eqref{csactionads} we obtain,
\begin{align}\label{adsaction}
	I_{\text{\tiny RAdS}}[\Phi,\zeta]=-\frac{k}{2\pi}\int\extd u \extd x\,\Big [ 2\frac{V'}{\ell}+2Y'-\frac{1}{2}(\dot \Phi\zeta'+\dot \zeta \Phi')+a\Big ]\,, 
\end{align}
again, we assume the fields are zero or have the same value at $x=\pm\infty$, so the first two terms can be ignored and thus we get $I_{\text{\tiny RAdS}}=I_{\text{\tiny RFlat}}+\frac{k}{2\pi\ell}\int\extd^2x \,\dot\zeta$, where $I_{\text{\tiny RFlat}}$ is given in \eqref{wzwf}.

To write the boundary action in terms of the 2D diffeomorphism fields $(f,\alpha)$, again we pay attention to the transformation of the state-dependent functions $(a,\eta)$ which in the AdS case are the following \cite{Afshar:2015wjm}
\begin{align}
    \delta a=\dot\epsilon\, a+\epsilon\,\dot a-\ddot\epsilon-2\dot\alpha/\ell\,,\qquad
    \delta T=2\dot\epsilon\, a+\epsilon\,\dot  a+\dot\alpha \,a+\ddot\alpha\,,
\end{align}
with their finite form being,
\begin{align}\label{adsfinitetrans}
\tilde a\circ f&=\frac{1}{\dot f}a+\frac{\ddot f}{\dot f^2}+\frac{2}{\ell}\alpha'\circ f\,,\nn\\
\tilde T\circ f&=\frac{1}{\dot f^2}T-\frac{1}{\dot f}a\,\alpha'\circ f-\frac{\ddot f}{\dot f^2}\alpha'\circ f-\alpha''\circ f-\frac{1}{\ell}(\alpha'\circ f)^2\,,
\end{align} 
where $T=\dot\eta+a\,\eta$. In the $\ell\to \infty$ limit these transformation laws reduce to \eqref{finitetrans}. Here, in the finite $\ell$ case, we do not need the explicit form of the transformation of the $\eta$ field since the covariant combination that appears in the calculation is always $T$. In fact, matching \eqref{grgfads} with the gauge field along $L_0$ and $L_{-1}$ after implementing \eqref{constrainteqs} leads to
\begin{align}\label{ads1steq}
	a=\dot \Phi-\frac{\dot\zeta}{\ell}\,,\qquad T=\ddot\zeta+\dot\zeta\,\dot\Phi\,.
\end{align}
The constant representative orbits can be obtained from \eqref{adsfinitetrans} using the inverse map $h=f^{-1}$
\begin{align}\label{a0h1}
	a&=\dot h\,a_0-\frac{\ddot h}{\dot h}+\frac{2}{\ell}\dot\alpha\,,\\
	T&=	\dot h^2\,T_0 - a_0\,\dot\alpha\,\dot h+\frac{\ddot h}{\dot h}\dot \alpha-\ddot \alpha-\frac{1}{\ell}\dot\alpha^2\,.\label{T0h2}
\end{align}
These equations determine the value of the function $(a,\eta)$ in the phase space by the action of the group of asymptotic symmetries on the constant representative orbit with $(a_0,\eta_0)$.
If we combine these two equations with \eqref{adsfinitetrans} we are led to a differential equation for $\zeta$,
\begin{align}
T_0\,\dot h^2+\left(\frac{\ddot h}{\dot h}-a_0\,\dot h\right)(\dot\zeta+\dot\alpha)-(\ddot\zeta+\ddot\alpha)-\frac{1}{\ell}(\dot\zeta+\dot\alpha)^2=0\,,
\end{align}
where $T_0$ is the zero mode of $T$.
This equation together with the first equation in \eqref{ads1steq} have the following solution for $\zeta$ and $\Phi$ in terms of $(h,\alpha)$,\footnote{The most general solution of course contains integration constants which we set to zero here.}
\begin{align}\label{fiedredef1}
	\zeta&=-\alpha -\frac{\ell}{2}  \left({a_0}+\sqrt{{a_0}^2+{4 {T_0}}/{\ell}}\right)h \,,\\
\Phi&=-\log\dot h+\frac{1}{\ell}\alpha+\frac{1}{2}  \left({a_0}-\sqrt{{a_0}^2+{4 {T_0}}/{\ell}}\right)h \,.\label{fiedredef2}
\end{align}
We are now ready to use \eqref{fiedredef1} and \eqref{fiedredef2} in the action \eqref{adsaction} and find its final form in terms of the diffeomorphism fields $(h,\alpha)$,
\begin{align}\label{AdSRaction}
	I[h,\alpha]=\frac{k}{4\pi}\int\extd u\extd x\Big[&2T_0\,\dot h h'-a_0\left(\dot h\alpha'+h'\dot\alpha\right)+\dot\alpha\frac{\dot h'}{\dot h}+\alpha'\frac{\ddot h}{\dot h}-2a_0 \dot h+2\frac{\ddot h}{\dot h}-\frac{4}{\ell}\dot\alpha\nn\\&+\frac{\ell}{2}\left({a_0}+\sqrt{{a_0}^2+{4 {T_0}}/{\ell}}\right)
	\left(\frac{\ddot h\,h'}{\dot h}+\dot h'\right)-\frac{2}{\ell}\dot \alpha\alpha'
	\Big]
\end{align}
which reduces to the action \eqref{boundaryaction} in the flat limit $\ell\to \infty$. This action can be considered as a two-dimensional warped-Schwarzian theory when the U(1) Kac-Moody level is present. The field equations corresponding to variation of $\alpha$ and $h$ are the following
\begin{align}\label{eom23}
    &\partial_u \left(a_0 h-\log\dot h+2\frac{\alpha}{\ell}\right)'=0\,,\\
    &\partial_u\left(a_0\alpha'+\frac{\dot\alpha'}{\dot h}-2T_0h'\right)=0\,.
\end{align}
One can verify that equations \eqref{eom23} reproduce equations \eqref{eom67} ($a'=0=T'$) upon using \eqref{a0h1}.

\section{Euclidean theory}\label{sec5}

Above we revisited the fact that 
if we consider the theory in periodic ‌Bondi time the 
phase space has warped conformal symmetry \cite{Afshar:2015wjm} and even constructed the corresponding boundary effective action in terms of a WZW theory for the gauge group ISL$(2,\mathbb R)$. In this section, we consider the theory at finite temperature $T_{\text{\tiny R}}=1/{\beta}$ by performing a Wick rotation $u \to iu \equiv \tau$ and working in Euclidean periodic Bondi time $\tau \sim \tau + \beta$.  All results of the last section apply here after a suitable analytic continuation;
\begin{align}\label{anltcont}
&u\to -i\tau\,,\qquad f(u)\to-if(\tau)\,,\qquad\alpha(u)\to\alpha(\tau)\,,\nn\\&
a(u)\to i a(\tau)\,,\qquad \eta(u)\to i\eta(\tau) \,,\qquad T(u)\to -T(\tau)\,,\qquad
\end{align}
and that  $f(\tau+\beta)=f(\tau)+\beta$. As an example the form of 
the gauge field $A_\tau\equiv -i A_u$ using the analytic continuation \eqref{anltcont} is
\begin{align}\label{Atau}
A_\tau&=a L_0-\frac{i}{2}TL_{-1}-iM_1-\frac{i}{\ell}L_1\,.
\end{align}
We will use the Euclidean $A_\tau$ for holonomy considerations in the following section for defining a black hole in the theory through its properties, such as regularity at the horizon in Euclidean spacetime;
\begin{figure}[h]
\centering
\includegraphics[width=4cm]{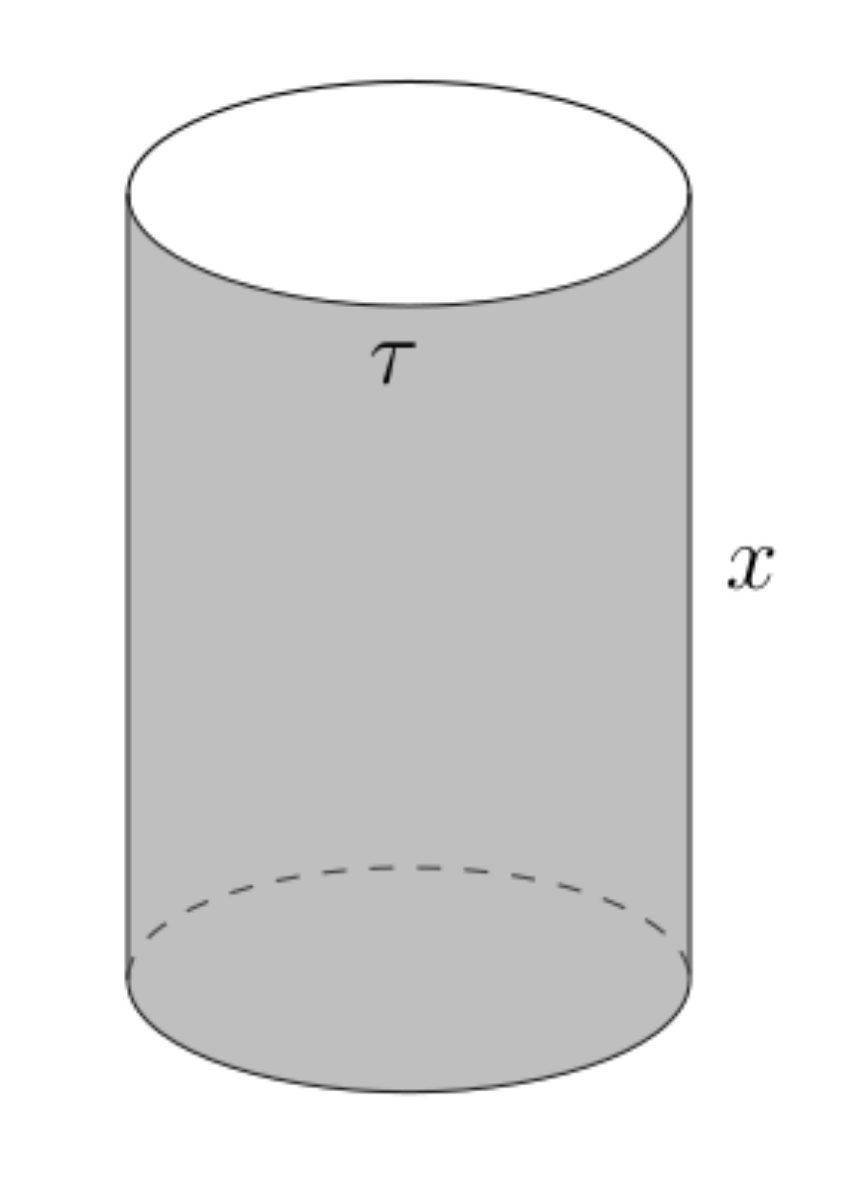}
\end{figure}
 
\subsection{Holonomy condition}
Our background has a horizon; thus, we should impose horizon regularity conditions in Euclidean signature. Translating this condition for the Euclidean gauge connection $A_\tau$ for the boosted Rindler AdS$_3$ \eqref{Atau} or the boosted Rindler flat ($\ell\to\infty$),  amounts to imposing the triviality of the holonomy of the Euclidean connection along contractible cycles \cite{Gutperle:2011kf, Ammon:2011nk, Kraus:2011ds} which is $\tau$ in our case. This necessitates the holonomy as a group element belonging to the center of the group \cite{Bunster:2014mua};
\begin{align}\label{Holcond}
    \text{Hol}[A_\tau]=\exp\oint A_{\tau}\in Z[G]\,,
\end{align}
where $\oint\equiv\int_0^\beta\extd\tau$. However, this condition is insufficient as we should consider the global properties of the actual gauge group.
The actual gauge group in the AdS$_3$ case (negative cosmological constant $\Lambda=-1/\ell^2$)  is 
\begin{align}
    \text{SO}(2,2)= \text{Spin}(2,2)/{\mathbb Z}_2\,,
\end{align}
where 
 ${\mathbb Z}_2$ is the centre of Spin$(2,2)\simeq \text{SL}(2,\mathbb R)\times\text{SL}(2,\mathbb R)$. 
Thus both elements $\{\mathds{1},-\mathds{1}\}\in {\mathbb Z}_2$ are trivial elements in the actual SO$(2,2)$ group. 
This amounts to consider the holonomy Hol[$A_\tau$]  in \eqref{Holcond}  to be trivial in SO$(2,2)$ but nontrivial in Spin$(2,2)$ as it corresponds to minus identity. In other words, we require the configuration \eqref{bcnhfo} to be regular (single-valued) as a gauge connection in SO$(2,2)$ group but not in Spin$(2,2)$ \cite{Castro:2011iw, Cotler:2018zff,Cotler:2020ugk}. 

The gauge group in the $\ell\to\infty$ case, sometimes referred to as ISL$(2,\mathbb R)$, is the double cover of the Poincar\'e group; ISO$(2,1)$ = ISL$(2,\mathbb R)/{\mathbb Z}_2$. 
As a consequence, the center of the 3D Poincar\'e group is given by,
\begin{align}
	Z(\text{ISO}(2,1))=Z(\text{ISL}(2,\mathbb R))/{\mathbb Z}_2\,.
\end{align}
Accordingly, the holonomy is a center element in both ISO$(2,1)$ and ISL$(2,\mathbb R)$ but by requiring the connection to be regular only in the former and not in the latter; Hol$[A_\tau]=-\mathds{1}$. In fact for the   Rindler space which is the vacuum of our theory with $(a_0,\eta_0)=(\frac{2\pi}{\beta},0)$ the holonomy of the Euclidean connection is $-\mathds{1}$ as we notice below.

\paragraph{Flat Rindler holonomy condition.}
In order to evaluate the holonomy condition \eqref{Holcond} for the gauge field \eqref{Atau}, we note that the group element $H=e^{B}$ belongs to the center of the group $G$ iff for any arbitrary member of the group $g=e^{X}$ we have Ad$_H X=X$ \cite{,Afshar:2020dth}.\footnote{
	Implying 
	\begin{align}
		[B,X]+\frac{1}{2}[B,[B,X]]+\frac{1}{3!}[B,[B,[B,X]]]+...=0   \,.
	\end{align}
}
If we apply this condition to the Euclidean gauge field $A_\tau$ in the flat limit $\ell\to \infty$ (at fixed $x$) we find the following conditions
\begin{align}\label{holc}
\oint a(\tau)=\pm2\pi i N\qquad (N\in\text{odd numbers}) \,,\qquad \oint T =\oint a\,\eta=0\,.
\end{align}
The first condition in \eqref{holc} always fixes the zero mode of the free function $a(\tau)$ as $a_0^{\text{\tiny E}}=\pm\frac{2\pi i N}{\beta}$. This means that the zero mode of $a(u)$ in the metric \eqref{brbc} should be $a_0=\mp\frac{2\pi N}{\beta}$ since $a(u)=i a(\tau)$. On the other hand, we always require $a(u)>0$ to guarantee that $u$ is a time-like direction. This is ensured only if we choose the lower sign in \eqref{holc}  and thus $\oint a(\tau)=-2\pi i N$ where $N$ is an odd number. Imposing the second condition to the boosted flat Rindler vacuum specified by $(a(\tau)=-\frac{2\pi i N}{\beta},\eta(\tau))$ fixes $\eta=\dot\sigma$ for a periodic function $\sigma$ which means $\eta_0=0$. For a general background specified by $(a(\tau),\eta(\tau))$ after 
Fourier 
expanding  
we have
\begin{align}\label{2ndHolcon}
    T_0^{\text{\tiny E}}=-\frac{2\pi i N}{\beta} \eta_0^{\text{\tiny E}}+\sum_{n\neq0}a_n\eta_{-n}=0\,,\quad N\in \text{ odd numbers}\,. 
\end{align}

This has a crucial consequence for the field $\eta$ which appears in the metric. It simply says that $\eta$ has no zero mode in $\tau$ unless higher modes of $a$ and $\eta$ are turned on. So the vacuum of the theory satisfying the trivial holonomy condition is Rindler space with $\eta_0=0$.\footnote{In this case the vacuum has four maximal global isometries compatible with asymptotic symmetries \cite{Afshar:2015wjm}.} The appearance of the boosted Rindler space satisfying the holonomy condition relies on turning on the higher modes. In our theory, this implies that higher mode black hole solution with non-zero $\eta$-term should satisfy \eqref{2ndHolcon}.
One can also translate the second trivial holonomy condition \eqref{holc} to an integral equation between the $x$-supertranslation mode $\alpha$ and the $\tau$-diffeomorphism $h$ using equation \eqref{T0h2}
	\begin{align}\label{hol45}
		-\oint \dot\alpha\Big(\frac{2\pi i N}{\beta}\dot h+\frac{\ddot h}{\dot h}\Big)=0\,.
	\end{align}
This condition is completely off-shell and fixes the allowed supertranslations in terms of Diff$(S^1)$ transformations such that the boundary preserving allowed diffeomorphisms also maintain the background as a black hole. 
Once we have their field equations using the action functional, these functions become more confined on-shell.

\paragraph{Rindler AdS$_3$ Holonomy condition.}
In the finite $\ell$ case corresponding to the boosted Rindler AdS$_3$ configuration  \eqref{rads}, evaluating the center of the group, independent of any chosen representation is hard. Since the algebra is semi-simple, we choose a representation and evaluate the holonomy accordingly. The fundamental representation of the so$(2,2)$ algebra \eqref{so22algebra} consists of six $4\times 4$ matrices as\footnote{We notice that the non-degenerate metric using the ordinary trace in this representation does not correspond to the bilinear form below \eqref{so22algebra}, in fact it reproduces the exotic bilinear form \cite{Witten:1988hc}. However, since we do not use any trace in calculating the holonomy, it is safe to use this representation for our purpose.}
\begin{align}\label{so22rep}
	M_0&=\frac{1}{2\ell}\begin{pmatrix}
		-1& 0 & 0 &  0\\
		0& 1& 0 & 0 \\
		0 & 0 &1 & 0 \\
		0&0 & 0 &-1 \\
	\end{pmatrix}\,,\quad M_{-1}=\frac{1}{\ell}\begin{pmatrix}
		0& 0 & 0 &  0\\
		1& 0& 0 & 0 \\
		0 & 0 & 0 & 0 \\
		0&0 &1 & 0 \\
	\end{pmatrix}\,,\quad M_{1}=\frac{1}{\ell}\begin{pmatrix}
		0& -1 & 0 & 0\\
		0& 0& 0 & 0 \\
		0 & 0 & 0 & -1\\
		0&0 & 0 & 0 \\
	\end{pmatrix}\,,\nn\\\nn\\ L_0&=\frac12\begin{pmatrix}
		-1& 0 & 0 &  0\\
		0&1& 0 & 0 \\
		0 & 0 &-1&0 \\
		0&0  & 0 &1 \\
	\end{pmatrix}\,,\qquad L_{-1}=\begin{pmatrix}
		0& 0 & 0 &  0\\
		1& 0& 0 & 0 \\
		0 & 0 & 0 & 0 \\
		0&0  & -1 & 0 \\
	\end{pmatrix}\,,\,\quad L_{1}=\begin{pmatrix}
		0&-1 & 0 &  0\\
		0& 0& 0 & 0 \\
		0 & 0 & 0 &1 \\
		0&0  & 0 & 0 \\
	\end{pmatrix}\,.
\end{align}
The holonomy of the bulk Chern-Simons Euclidean connection $A_\tau$ given in \eqref{Atau} using the representation \eqref{so22rep} is the following
\begin{align}
	\text{Hol}[A_\tau]&=
	\left(
	\begin{array}{cccc}
		\cosh \left(\frac{S}{2 }\right)-\frac{ \beta a_0^{\text{\tiny E}} \sinh \left(\frac{S}{2 }\right)}{S} & \frac{4i \beta/\ell  \sinh \left(\frac{S}{2 }\right)}{ S} & 0 & 0 \\
		-\frac{i\beta T_0^{\text{\tiny E}} \sinh \left(\frac{S}{2 }\right)}{S} & \frac{ \beta a_0^{\text{\tiny E}} \sinh \left(\frac{S}{2 }\right)}{S}+\cosh \left(\frac{S}{2 }\right) & 0 & 0 \\
		0 & 0 & e^{-\frac{\beta a_0^{\text{\tiny E}}}{2}} & 0 \\
		0 & 0 & \frac{ iT_0^{\text{\tiny E}} \sinh \left(\frac{\beta a_0^{\text{\tiny E}}}{2}\right)}{ a_0^{\text{\tiny E}}} & e^{\frac{\beta a_0^{\text{\tiny E}}}{2}} \\
	\end{array}
	\right)
\end{align}
where  
$S=\beta\sqrt{ (a_0^{\text{\tiny E}})^2+4 T_0^{\text{\tiny E}}/\ell}$. 
To have trivial holonomy corresponding to regular metrics we need to impose Hol$[A_\tau]=-\mathds{1}$ which is satisfied if
$
\beta a_0^{\text{\tiny E}}=\pm2\pi i N_1$ and $S= \pm2\pi i N_2$ for $N_1,N_2$ being odd numbers. Again similar to the flat case,  the fact that $u$ in the metric \eqref{brbc} remains time-like suggests picking the lower sign for $a_0^{\text{\tiny E}}$. If we rewrite these conditions on zero modes we have
\begin{align}\label{a0hol}
a_0^{\text{\tiny E}}&=-\frac{2\pi i N_1}{\beta}\,,\qquad N_1=1,3,5\cdots\,,\\	T_0^{\text{\tiny E}}&= \frac{\pi^2\ell}{\beta^2} ( N_2^2-N_1^2)=
	\pm \frac{4\pi^2\ell}{\beta^2}N_3\,,\qquad N_3=0,2,4,\cdots\,,\label{T0hol}
\end{align}
implying that the allowed zero modes $a_0$ and $T_0$ are quantized in odd and even integers respectively in the boosted Rindler AdS case. Unlike the case of flat Rindler, the zero mode of the function $T$ can be non-zero. Again we may represent the last holonomy condition on the zero mode of $T$ in terms of a constraint integral equation among the allowed 2D diffeomorphisms. Using equation \eqref{a0h1} and \eqref{T0h2} the first trivial holonomy condition \eqref{a0hol} is automatically satisfied, while the second holonomy condition \eqref{T0hol} gives the off-shell constraint
\begin{align}
      \pm \frac{4\pi^2\ell}{\beta}N_3=\oint\Big(\pm \frac{4\pi^2\ell}{\beta^2}N_3\dot h^2\, + \frac{2\pi i N_1}{\beta}\dot\alpha\,\dot h+\frac{\ddot h}{\dot h}\dot \alpha-\frac{1}{\ell}\dot\alpha^2\Big)\,.
\end{align}

\subsection{Flat Rindler boundary action}
The presence of the warped conformal symmetry in the (Rindler) near horizon spacetime advances the proposal that the effective action depicting the near horizon (non-extremal) behavior should be a warped-Schwarzian theory \cite{Afshar:2019tvp}. 

In the boosted Rindler flat case ($\ell\to\infty$), the analytic continuation of the boundary action \eqref{boundaryaction} after imposing the regularity condition at the horizon namely $a_0^{\text{\tiny E}}=-\frac{2\pi i N}{\beta}$, $T_0=0$, leads to
\begin{align}\label{baction23}
	I^{\text{\tiny Rflat}}[h,\alpha]=
		\frac{k}{4\pi}\int \extd \tau\extd x \Big(\frac{2\pi i N}{\beta}(\dot h\alpha'+\dot\alpha h')+\dot\alpha\frac{\dot h'}{\dot h}+\alpha'\frac{\ddot h}{\dot h}+\frac{4\pi i N}{\beta} \dot h+2\frac{\ddot h}{\dot h}\Big)\,,
\end{align}
where $\cdot$ is the derivative w.r.t. the Euclidean Bondi time $\tau$ and ${}^\prime$ is the derivative w.r.t. the coordinate $x$. 
By assuming that both $\alpha$ and $h$ fields have the same value or are zero at boundaries at $x=\pm\infty$, we can simplify the action using partial integration,
\begin{align}\label{flatEuclaction}
 I^{\text{\tiny Rflat}}[h,\alpha]=
		\frac{k}{2\pi}\int\extd \tau\extd x \Big(-\alpha\big(\frac{2\pi i N}{\beta} \dot h+\frac{\ddot h}{\dot h}\big)'+\frac{2\pi i N}{\beta} \dot h+\frac{\ddot h}{\dot h}\Big)\,.  
\end{align}
The variation of the action w.r.t $\alpha$ yields the equation of motion 
\begin{align}\label{al}
   &\partial_\tau\left(-\frac{2\pi iN}{\beta}h-\log\dot h\right)'=0\quad\to\quad  -\frac{2\pi iN}{\beta}h-\log\dot h= C_1(\tau)+C_2(x)\,.
\end{align}
We remind that this solution should also satisfy the first holonomy condition in \eqref{holc} which leads to 
\begin{align}
    \oint \dot C_1(\tau)=- 2\pi i N\;\to\; C_1(\tau+\beta)-C_1(\tau)=- 2\pi i N
\end{align}
where $C_1(\tau)$ in a most simple form can be solved as $C_1(\tau)=-\frac{2\pi i N}{\beta}\tau$.
We can also solve the function $\alpha$ using the equation derived by varying the action w.r.t. $h$,
\begin{align}\label{eh}
   \partial_\tau\left(-\frac{2\pi i N}{\beta}\alpha'+\frac{\dot\alpha'}{\dot h}\right)=0\quad\to\qquad \log\dot\alpha'=\frac{2\pi i N}{\beta}\tau+C_3(x)\,,\end{align}
where we used the solution for $\dot h$ in the last equality.

\subsection{Rindler AdS$_3$ boundary action} In this case, the  holonomy condition $\text{Hol}[A_{\tau}]=-{\mathds{1}}$  resulted in the zero mode for $a$ as in  \eqref{a0hol} which is the same as in the flat case \eqref{holc}, while the zero mode for $T$ has changed according to \eqref{T0hol}. As a consequence of applying these background values,
 and after analytically continuing the boundary action \eqref{AdSRaction} we have
\begin{align}\label{adsbndraction}
I^{\text{\tiny RAdS}}[h,\alpha]=I^{\text{\tiny Rflat}}[h,\alpha]+\frac{k}{2\pi}\int \extd \tau\extd x \Big[\pm \frac{4\pi^2\ell}{\beta^2}N_3\dot h h'\pm\frac{\pi i\ell  N_4}{2\beta}
\left(\frac{\ddot h\,h'}{\dot h}+\dot h'\right)-\frac{\dot \alpha\alpha'}{\ell}-\frac{2\dot\alpha}{\ell}\Big]\,,
\end{align}
where $ N_4=N_1+N_2$ and $N_3=\pm(N_1^2-N_2^2)/4$ are both even numbers. 
\section{Discussion}

We worked out the boundary effective action governing the dynamics of 2D allowed diffeomorphisms corresponding to the field theory of the boundary gravitons on the 3D Rindler spacetime. Since our theory always involves a horizon, in the Euclidean theory we imposed the regularity condition at the horizon via imposing triviality of the holonomy conditions. This fixed the zero modes of state-dependent functions $(a,T)$ corresponding to the exceptional coadjoint orbits of the asymptotic symmetry group, the warped Virasoro group.  
\paragraph{Entropy.} The entropy density $s=S/\Delta$ with $\Delta=\int_{-\infty}^{\infty}\extd x$ in the canonical ensemble can be calculated using the on-shell Euclidean action, which in this case using \eqref{flatEuclaction} and introducing $I^{\text{\tiny E}}\equiv iI_{\text{\tiny Wick-rot.}}$ leads to\footnote{This convention guarantees that the real contour for the Lagrange multiplier $\alpha$ leads to a well-defined Euclidean path integral.
 }
\begin{align}
    S=-I^{\text{\tiny E}}_{\text{\tiny on-shell}}=\frac{-ik}{2\pi}\int_{-\infty}^{\infty}\extd x\int_0^\beta\extd\tau\Big(\frac{2\pi i N}{\beta} \dot h+\frac{\ddot h}{\dot h}\Big)=\frac{-ik\Delta}{2\pi}\left(2\pi i N\right)=\frac{N\Delta}{4G}\,,
\end{align}
where in the last identity we used that $k=1/(4G)$. We note that $N$ is an odd number with $N=1$ for the Rindler background leading to $s=1/(4G)$. For Rindler thermodynamics see  \cite{Laflamme:1987ec} and appendix B of \cite{Afshar:2015wjm}.

Recent advancements in AdS$_2$/CFT$_1$ correspondence came with the discovery that Euclidean JT gravity has a dual description of an ensemble-averaged matrix model \cite{Saad:2019lba}. This brought up the idea that certain states in the bulk have ensemble averaging holographic interpretation \cite{Schlenker:2022dyo}. There have been efforts to check if this is a unique feature of JT gravity or if it can be generic in higher dimensional examples of AdS/CFT as well \cite{Cotler:2020ugk,Pollack:2020gfa}. One of the major statements in \cite{Schlenker:2022dyo} to address this feature is that an observable that does not involve black holes sees no sign of ensemble averaging. In other words, ensemble averaging only affects observables that do involve black holes.  In fact 
the Euclidean JT gravity's special random matrix feature was raised by considering  AdS$_2$ Rindler boundary conditions. It is thus tempting to consider the Euclidean 3D Einstein gravity with Rindler boundary conditions as a more realistic theory of gravity and test the ensemble interpretation of the boundary theory in this case. 

3D gravity and 2D dilaton-gravity are related via the dimensional reduction \cite{Achucarro:1993fd}. 
The two famous dilaton-gravity theories exhibiting a BF-type formulation are the Jackiw-Teitelboim and the Cangemi-Jackiw gravity theories whose boundary effective action has shown to be the so-called 1D Schwarzian and warped-Schwarzian theories. It is tempting to investigate the same dimensional reduction at the level of the 2D boundary action obtained in this work and the 1D (warped)-Schwarzian ones as suggested by \cite{Mertens:2018fds}. 
\section*{Acknowledgment} 
We thank the anonymous referee for his/her valuable comments. We thank Daniel Grumiller for his comments on the draft. 
NA would like to thank Monica Guica, Pavel Putrov and 
Cumrun Vafa for 
discussions on general aspects of holography and the Hamiltonian reduction in 3D gravity during her participation in the Spring School on Superstring Theory and Related Topics 2023, the New Pathways in Exploration of Quantum Field Theory 2023, and Quantum Gravity beyond Supersymmetry III and String-Math 2024 at ICTP. She also appreciates the Abdus Salam International Centre for Theoretical Physics (ICTP) for providing financial support throughout her stay. This research was supported by the Iran National Science Foundation (INSF), project No. 4000132. 
\bibliographystyle{fullsort.bst}
 
\bibliography{references} 

\end{document}